\documentclass[final,5p,times,twocolumn,number]{elsarticle}

\usepackage{amssymb}
\usepackage{graphicx}
\usepackage{hyperref}
\usepackage{hypernat}
\usepackage{float}

\journal{Chemical Physics Letters}
\begin{document}

\begin{frontmatter}

\title{\textit{Ab initio} potential curves for the X $^2 \Sigma_{u}^+$ and B $^2 \Sigma_{g}^+$ states of Be$_{2}^+$: \\ Existence of a double minimum}

\author[]{Sandipan Banerjee\corref{cor1}}
\ead{banerjee@phys.uconn.edu}

\author[]{Jason N. Byrd}

\author[]{Robin C\^ot\'e}

\author[]{H. Harvey Michels}

\author[]{John A. Montgomery, Jr.}

\cortext[cor1]{Principal corresponding author; Fax: +1 860 486 3346}

\address{Department of Physics - University of Connecticut, Storrs, CT 06269-3046, USA.}

\begin{abstract}
We report \textit{ab initio} calculations of the X $^2 \Sigma_{u}^+ $ and B $^2 \Sigma_{g}^+$ states of the Be$_{2}^+$ dimer. Full valence configuration interaction calculations were performed using the aug-cc-pVnZ basis sets and the results were extrapolated to the CBS limit. Core-core, core-valence effects are included at the CCSDT/MTsmall level of theory. Two local minima, separated by a large barrier, are found in the expected repulsive B $^2 \Sigma_{g}^+$ state. Spectroscopic constants have been calculated and good agreement is found with the recent measurements of Merritt {\em et al.} Bound vibrational levels, transition moments and lifetimes have also been calculated.\\
\end{abstract}


\end{frontmatter}

\section*{Introduction}

In recent years, there has been a lot of interest in ultracold atom-ion scattering \cite{zhang2009} in the atomic and molecular physics community. The experimental realization of Bose-Einstein condensation (BEC) has led to numerous applications involving charged atomic and molecular species. The cooling and trapping \cite{Weiner1999} of such charged gases at sub-kelvin temperatures is a topic of growing interest. The phenomena of charge transport like resonant charge transfer \cite{Robin_Dalgarno2000} and charge mobility \cite{Robin2000} at ultracold temperatures have also been studied in detail. Other emerging fields of interest include ultracold plasmas \cite{Plasma1}, ultracold Rydberg gases \cite{Rydberg1} and systems involving ions in a BEC \cite{Robin2002, BEC1}. \\

It is well known that the Be$_2$ dimer is a difficult problem for
computational quantum chemistry, due to the $2s-2p$ near degeneracy
in the beryllium atom\cite{Martin1999}.
This near-degeneracy problem also
arises in the less studied Be$_2^+$ dimer.  The ground X $^2\Sigma_{u}^+$
state is well described by a $1\sigma_g^2 1\sigma_u^2 2\sigma_g^2 2\sigma_u^{\phantom{2}}$
reference, but the B $^2\Sigma_{g}^+$ state has a multi-reference character,
as previously discussed by Fischer {\em et al.} \cite{Fischer1991}. In this paper we present calculations on the Be$_{2}^+$ dimer that should be useful as a starting point for further studies in ultracold atomic and molecular physics. \\

We begin by describing the methods used in our calculations followed by a discussion of the
results which include the potential curves, spectroscopic constants, dipole moments,
lifetimes of the bound vibrational levels of the $^2\Sigma_{g}^+$ state and the analysis of long range behavior and determination of the Van der Waals coefficients.
The  B $ ^2\Sigma_{g}^+ $ state of Be$_{2}^+$ has a double minima instead of a purely repulsive
nature as one would expect. We have characterized both the inner (deep) and outer (shallow) well. \\

\begin{figure}[t]
 \centering
\includegraphics[width=1.0\linewidth, height=2.5in]{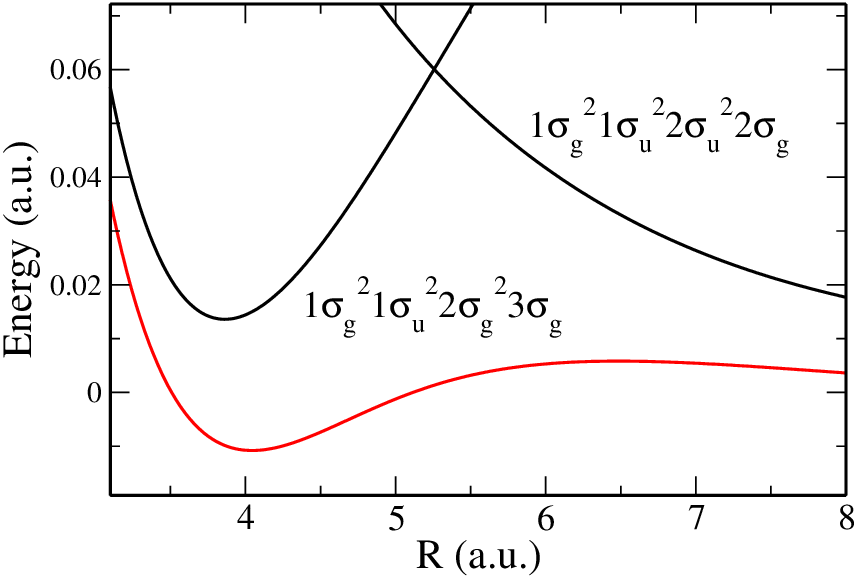}
\caption{[COLOR ONLINE] Dominant molecular orbital configurations in the B $^2\Sigma_{g}^+$ state. The curves in black shows the SCF curve crossings due to the change of the occupied molecular orbitals, whereas the curve in red shows the calculated FCI potential curve for the B $^2\Sigma_{g}^+$ state.}
\label{fig:scf_cross}
\end{figure}

\section*{Methods}

\begin{figure}
 \centering
\includegraphics[width=1.0\linewidth]{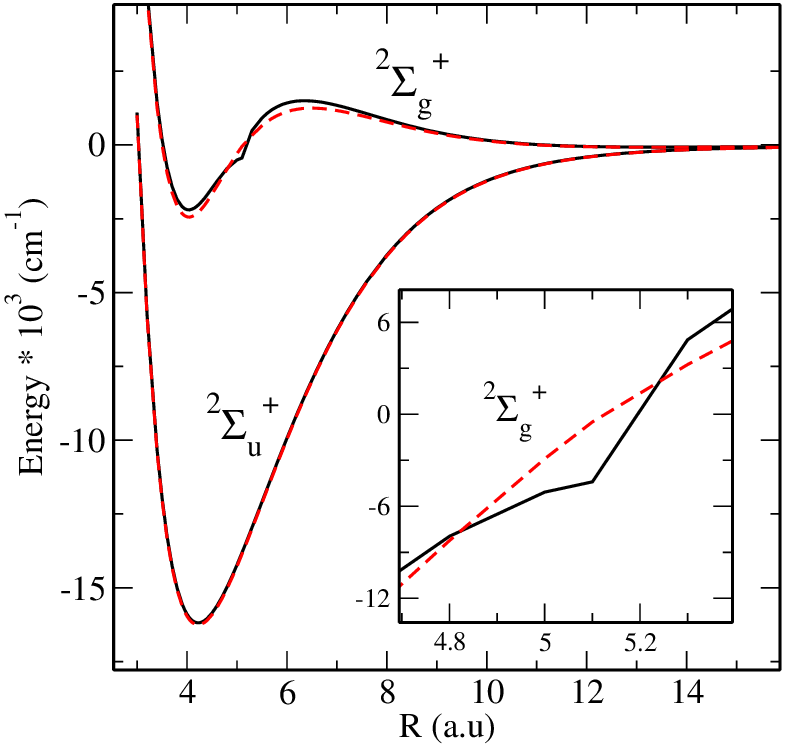}
\caption{[COLOR ONLINE] The curves in black show a CCSD(T) calculation done with aug-cc-pV5Z basis set, whereas the curves in red are a full CI calculation using the same basis set. The inset shows a discontinuity (black line) in the CCSD(T) curve for the B $^2\Sigma_{g}^+$ state due to the change in reference configuration at larger internuclear separation.}
\label{fig:ccsd_t}
\end{figure}

As mentioned above, the B $^2\Sigma_{g}^+$ state cannot be adequately described
by a single molecular orbital configuration.
At short internuclear separation, the dominant configuration is
$1\sigma_g^2 1\sigma_u^2 2\sigma_g^2 3\sigma_g^{\phantom{2}}$
while at large separation it becomes
$1\sigma_g^2 1\sigma_u^2 2\sigma_u^2 2\sigma_g^{\phantom{2}}$.
This behavior is shown in Fig. [\ref{fig:scf_cross}].
Preliminary calculations done at the CCSD(T) level of theory
find a kink in the potential curve for the B $^2\Sigma_{g}^+$ state
at the SCF curve crossing (see Fig. [\ref{fig:ccsd_t}]).  Valence
full configuration interaction (FCI) calculations were found to give a smooth
potential curve. \\

Therefore,
our computational approach is to perform valence FCI using
the augmented correlation consistent polarized valence n-tuple zeta (aug-cc-pVnZ) basis set of
Dunning.\cite{Dunning1989}
We then extrapolated the results from the aug-cc-pVQZ and aug-cc-pV5Z
\footnote{The aug-cc-pV5Z basis was created by adding diffuse primitives with the following exponents
to the published cc-pV5Z basis: s 0.013777, p 0.007668, d 0.0772, f 0.01375, g 0.174, h 0.225.}
calculations to the complete basis set (CBS) limit.
We have used Schwenke's linear formula \cite{schwenke05} to extrapolate the SCF energies. For extrapolating
the FCI correlation energies we have used the following formula given by Helgaker \cite{Helgaker98}:

\begin{equation}
E^{\infty}_{X Y} = \frac{X^3 E_{X}^{corr} - Y^3 E_{Y}^{corr}}{X^3 - Y^3}\, ,
\end{equation}

where X, Y are 4, 5 corresponding to the aug-cc-pVQZ and aug-cc-pV5Z basis sets.
The total valence energy is the sum of the extrapolated SCF and full CI correlation energies.  
Core-core (CC) and core-valence (CV) correlations were calculated as the difference between
all-electron and frozen core
CCSDT \cite{CCSDT} calculations done with Martin's MTsmall basis set \cite{MTsmall1999}.
The MTsmall basis set consists of a completely uncontracted cc-pVTZ basis set augmented with
two tight d and one tight f functions. The calculated potential energy curves are corected for the
effects of basis set superposition error by the counterpoise method of Boys and Bernardi\cite{CPcorrection}.
The CBS extrapolation increased the well depths of the
X $^2\Sigma_{u}^+$ and B $^2\Sigma_{g}^+$ states
by $\sim$ 40 cm$^{-1}$, however D$_{0}$ for the outer well in the B $^2\Sigma_{g}^+$ state was unchanged.
Scalar relativistic corrections were estimated to be $\sim$ 10 cm$^{-1}$ and are neglected. \\

The FCI calculations were done using the \texttt{MOLPRO 2009.1} \cite{MOLPRO_brief}
and \texttt{PSI3} \cite{PSI3} electronic structure programs running on a Linux workstation (2 quad core Intel Xeon E5520 CPU).
The core-core and core-valence corrections were done with the multi-reference coupled cluster (\texttt{MRCC})
program \cite{MRCC} of M. K\'{a}llay.

Le Roy's \texttt{LEVEL} program \cite{LEVEL} has been used to calculate the bound
vibrational levels, Frank-Condon factors and Einstein A coefficients. Using these Einstein A
coefficients we were able to calculate the lifetimes of all vibrational levels of the $ ^2\Sigma_{g}^+$ state.

\section*{Results and Discussions}
\subsection*{Potential Curves and Spectroscopic Constants}

\begin{figure}[t]
 \centering
\includegraphics[width=1.0\linewidth]{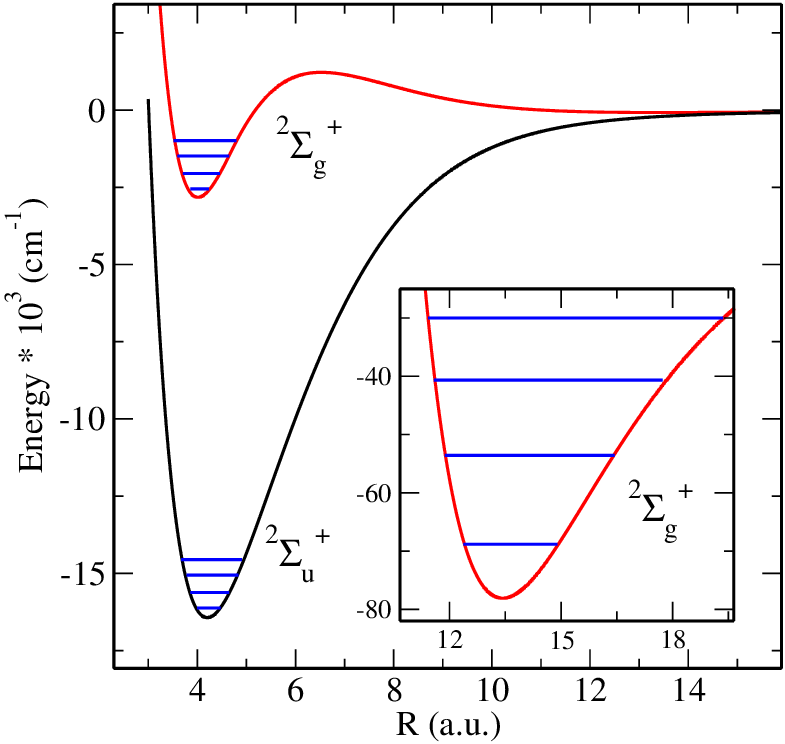}
\caption{[COLOR ONLINE] The figure shows the calculated \textit{ab initio} potential curves of Be$_{2}^+$. The inset is a magnification of the shallow long-range well in the B $^2\Sigma_{g}^+$ state (in red). The positions of the first four bound vibrational levels are shown (in blue) for both states of Be$_{2}^+$ dimer. Note that the energy scale for the inset is in cm$^{-1}$.}
\label{fig:curves}
\end{figure}

Fig. [\ref{fig:curves}] shows the \textit{ab initio} potential curves for the lowest $^2\Sigma_{u}^+ $
and $ ^2\Sigma_{g}^+ $ states of Be$_{2}^+ $ dimer. We have used a standard Dunham analysis \cite{Dunham1932}
to calculate the spectroscopic constants (Table \ref{tab:spec}).

The bond dissociation energy (D$_{0}$) is 16172 cm$^{-1}$ is in very good agreement with the
experimental data of the Be$_{2}^+$ dimer \cite{Heaven2008}. The X $^2\Sigma_{u}^+ $ state supports approximately 70 bound vibrational levels. \\ 

The B $^2\Sigma_{g}^+$ state which was expected to be repulsive has two minima instead.
Both these wells support bound vibrational states. The outer well has 12 bound levels which are long lived ($\sim$ ms).

\begin{table}[h!]
\centering
\caption{Calculated spectroscopic constants of Be$_{2}^+$}
\label{tab:spec}
\scalebox{0.8}{%
\begin{tabular}{c c c c c c}\\ \\
\hline \hline \\
State & r$_{e}$  ({\AA})  &  B$_{e}$ (cm$^{-1}) $ & $\omega_{e}$ (cm$^{-1})$ & $\omega_{e} x_{e}$ (cm$^{-1})$ & D$_{0}$ (cm$^{-1})$  \\ \\
\hline \hline \\
X$^2 \Sigma_{u}^+$  & 2.223 & 0.756 & 525.299 & 4.454 & 16172 \\ \\
Exp. \cite{Heaven2008} & & & 498(20) & & 16072(40) \\ \\
B$^2 \Sigma_{g}^+$ (In)  & 2.123 & 0.829 & 547.452 & 11.681 & 2550 \\ \\
B$^2 \Sigma_{g}^+$ (Out) & 7.106 & 0.074 & 33.703 & 3.548 & 69 \\ \\
\hline \hline
\end{tabular}}
\end{table}

\newpage
\subsection*{Transition Moments and Lifetimes}

\begin{figure}[t]
 \centering
\includegraphics[width=0.95\linewidth]{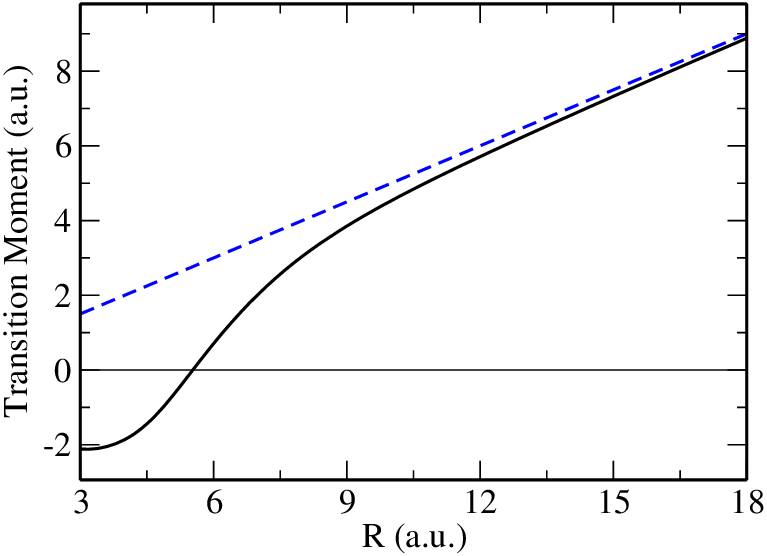}
\caption{[COLOR ONLINE] The figure shows a plot of the computed electronic dipole transition moment $\mu_{XB}$ coupling the B $^2\Sigma_{g}^+$ to the X $^2\Sigma_{u}^+$ state. The dotted line (in blue) shows the a plot of R/2. }
\label{fig:trm}
\end{figure}

To compute the transition moments coupling the X $^2 \Sigma_{u}^+ $ and B $^2 \Sigma_{g}^+$ states of Be$_{2}^+$ dimer, we have used a 16 orbital complete active space self consistent field (CASSCF) wavefunction as a reference for performing multi-reference configuration interaction (MRCI) calculations. The transition moment for electric dipole transitions is defined as,

\begin{equation}
\mu_{XB} (R) = \langle B\,|\, er\, |\,X \rangle \, ,
\end{equation} 

where $|X\rangle$ and $|B\rangle$ are the electronic wave functions corresponding to the states X $^2\Sigma_{u}^+$ and B $^2\Sigma_{g}^+$ when the two Be cores are separated by the distance R. Fig. [\ref{fig:trm}] shows a plot of the computed electronic dipole transition moment between the  B $^2\Sigma_{g}^+$ and the X $ ^2\Sigma_{u}^+$ ground states of Be$_{2}^+$. The transition moment $\mu_{XB}$ asymptotically follows the classical dipole behavior, $\mu_{XB}$ $\sim$ R/2. The curve shows a zero-crossing at around 5.5 bohrs which is approximately the same distance at which the dominant molecular orbital configuration changes from (1$\sigma_{g}^2$ 1$\sigma_{u}^2$ 2$\sigma_{g}^2$ 3$\sigma_{g}$) to (1$\sigma_{g}^2$ 1$\sigma_{u}^2$ 2$\sigma_{g}$ 2$\sigma_{u}^2$) in the B $^2\Sigma_{g}^+$ state of Be$_{2}^+$ (Fig. [\ref{fig:scf_cross}]). \\
 
The calculated potential curves and the electronic transition dipole moments were used as input to Le Roy's
\texttt{LEVEL} program to calculate the Einstein A coefficients coupling the vibrational bound levels of the 
B $^2\Sigma_{g}^+$ state to the X $^2\Sigma_{u}^+$ state. We have also calculated the radiative lifetimes
(Table: [\ref{tab:lifetimes}]) of the vibrational levels in the B $^2\Sigma_{g}^+$ state using these
Einstein A coefficients. Note that the bound levels in the shallow outer well are extremely long-lived
($\sim$ $10^{-3}$ s) in comparison to the levels in the inner well ($\sim$ $10^{-7}$ s). Our results for $v^\prime$ = 0 - 3 agree well with the results of Fischer {\em et al.} \cite{Fischer1991}.

\begin{table}[ht!]
\centering
\caption{Radiative lifetimes of the vibrational levels of the B $^2\Sigma_{g}^+$ state (in s).}
\label{tab:lifetimes}
\begin{tabular}{c c }\\ 
\hline \hline \\
$v^ \prime$ &  B $^2\Sigma_{g}^+$ $ \rightarrow $ X $^2\Sigma_{u}^+$  \\ \\ 
\hline \hline \\
0 & 0.849 $ \times $ $10^{-7}$ \\
1 & 0.937 $ \times $ $10^{-7}$ \\
2 & 1.032 $ \times $ $10^{-7}$ \\
3 & 1.161 $ \times $ $10^{-7}$ \\
4 & 1.423 $ \times $ $10^{-7}$ \\
5 & 1.539 $ \times $ $10^{-7}$ \\ 
6 & 2.870 $ \times $ $10^{-3}$ \\
7 & 1.861 $ \times $ $10^{-3}$ \\
8 & 1.586 $ \times $ $10^{-3}$ \\
9 & 1.557 $ \times $ $10^{-3}$ \\
10 & 1.992 $ \times $ $10^{-3}$ \\
11 & 1.840 $ \times $ $10^{-3}$ \\
12 & 2.340 $ \times $ $10^{-3}$ \\
13 & 3.382 $ \times $ $10^{-3}$ \\
14 & 5.277 $ \times $ $10^{-3}$ \\
15 & 10.067 $ \times $ $10^{-3}$ \\
16 & 25.718 $ \times $ $10^{-3}$ \\
17 & 97.361 $ \times $ $10^{-3}$ \\
\hline \hline
\end{tabular}
\end{table}

\subsection*{Long Range Coefficients}

For large internuclear separations, the standard long-range form of the intermolecular potential is: 

\begin{equation}
\label{eqn:v-lr}
V_{LR}(R) = V_{\infty} - \frac{(\alpha_{1}/2)}{R^4} - \frac{(\alpha_{2}/2 + C_{6})}{R^6} -  ...  \pm E_{exch}\, ,
\end{equation}
where $E_{exch}$ is the exchange energy contribution and $V_{\infty}$ is the energy of the atomic asymptote (which we have set to zero). Note that $\alpha_{1}$ is the static dipole polarizability, $\alpha_{2}$ is the quadrupole polarizability and C$_{6}$ is the dispersion coefficient. $E_{exch}$ is repulsive (plus sign in Eq. [\ref{eqn:v-lr}]) for the $ ^2\Sigma_{g}^+ $ state and attractive (minus sign in Eq. [\ref{eqn:v-lr}]) for the $ ^2\Sigma_{u}^+ $ state. \\

All the parameters in Eq. [\ref{eqn:v-lr}] are common for both the X $ ^2\Sigma_{u}^+ $ and B $ ^2\Sigma_{g}^+ $ states. Neglecting higher order terms in Eq. [\ref{eqn:v-lr}], if we add the potentials for both states, the exchange term cancels and we get,

\begin{equation}
\label{eqn:fit}
- \frac{(V_{g}+ V_{u})}{2} \times R^4 = (\alpha_{1}/2) +\frac{{(\alpha_{2}/2} + C_{6})}{R^2}\, .
\end{equation}

Table [\ref{tab:lr-coeff}] lists the values of the long-range coefficients that we obtained from fitting our long-range data to Eq.[\ref{eqn:fit}]. We get $\alpha_{1}$ = 38.12 a.u. which is in excellent agreement with previous results \cite{Komasa2001}. We have also calculated the quadrupole polarizability from a finite-field calculation using \texttt{MOLPRO} which gives $\alpha_{2}$ = 300.01 a.u. Thus from the fit we were able to extract the value of the dispersion coefficient C$_{6}$ = 124.22. This is in good agreement with unpublished results of Mitroy \cite{Mitroy}.

\begin{table}
\centering
\caption{Long Range Coefficients for both X$ ^2\Sigma_{u}^+ $ and B$ ^2\Sigma_{g}^+ $ states (in a.u.) 
\label{tab:lr-coeff}}
\begin{tabular}{c c c c}\\ 
\hline \hline  
& $\alpha_{1}$  & $\alpha_{2}$ & C$_{6}$ 
 \\ \hline \hline \\
 This work & 38.12 & 300.01 & 124.23 \\ 
 Previous \cite{Komasa2001} & 37.76 & 300.98 & \\ 
 Previous \cite{Mitroy} &  & & 119.99  
  \\ \hline \hline 
\end{tabular}
\end{table}

\section*{Concluding Remarks}

Accurate \textit{ab initio} calculations have been performed on the X $^2\Sigma_{u}^+$ and B $^2\Sigma_{g}^+$
states of the Be$_{2}^+$ dimer. Since the $ ^2\Sigma_{g}^+ $ state has a shallow well near 13.4 bohr,
it was necessary to include diffuse functions in the basis sets to describe the well
accurately. Large augmented basis sets of the Dunning correlation consistent series were thus chosen and the results were also
extrapolated to the complete basis set limit.
We have corrected our valence only FCI results for core-core and core-valence effects
using CCSDT calculations with both full and frozen core using Martin's MTsmall basis set. \\

Since the $^2\Sigma_{g}^+$ state has not been experimentally observed we were unable to compare our theoretical values for
dissociation energies or spectroscopic constants\cite{Heaven2008}.
However there are recent experimental results for the bond dissociation energy and $\omega_{e}$ for the
$ ^2\Sigma_{u}^+ $ state which are respectively 16072 $\pm$ 40 cm$^{-1}$ and 498 $\pm$ 20 cm$^{-1}$.
These values compare well with our calculated results of 16172 cm$^{-1}$ and 525 cm$^{-1}$ . \\ 

\section*{Acknowledgements}
This work has been supported in part by the U.S. Department of Energy Office of Basic Sciences and the National Science Foundation. We would also like to thank J. Mitroy for sharing his calculated dispersion coefficients.\\

\end{document}